# On possibility of realization NEUTRINO-4 experiment on search for oscillations of the reactor antineutrino into a sterile state

A.P. Serebrov[a*], A.K. Fomin[a], V.G. Zinoviev[a], V.G. Ivochkin[a], Yu.E. Loginov[a], G.A. Petrov[a], V.A. Solovey[a], A.V. Chernyi[a], O.M. Zherebtsov[a], R.M. Samoylov[a], V.P. Martemyanov[b], V.G. Tsinoev[b], V.G. Tarasenkov[b], V.I. Aleshin[b], A.L. Petelin[c], S.V. Pavlov[c], M.N. Svyatkin[c], A.L. Izhutov[c], S.A. Sazontov[c], D.K. Ryazanov[c], M.O. Gromov[c], V.V.Afanasiev[c], V.I. Rykalin[d]

[a]*Petersburg Nuclear Physics Institute, Gatchina, Leningrad region, 188300 Russia*
[b]*National Research Centre "Kurchatov Institute", Moscow, 123182 Russia*
[c]*State Scientific Centre - Research Institute of Atomic Reactors, Dimitrovgrad-10, Ulyanovsk region, 433510 Russia*
[d]*Institute for High Energy Physics, Protvino, Moscow region, 142281 Russia*
[*]*e-mail: serebrov@pnpi.spb.ru*

One has investigated possibility of performing NEUTRINO-4 experiment on search for reactor neutrino oscillations into a sterile state at research reactors. The simulated experiment has been conducted at 16 MW reactor WWR-M in PNPI with the purpose of implementing a full scale experiment with the help of 100 MW reactor SM-3 in RIAR. Background conditions for making such an experiment have been examined at both reactors. The conclusion has been made on possible implementation of a full scale experiment NEUTRINO-4 at the reactor SM-3 in RIAR.

At present possibility of existence of a sterile neutrino is being widely discussed. It is assumed that due to a reactor antineutrino transition into a sterile state one can observe both the oscillation effect at short distances from the reactor and deficiency of reactor antineutrino flux at large distances [1,2].

As early as 1967 B.M. Pontekorvo initially assumed [3,4] that neutrino transitions into a sterile state were likely to occur. Further development of the idea of neutron oscillations allowed describing neutron oscillations with the model of three neutrino generations. The picture of this phenomenon can be sufficiently well represented by the matrix of Pontecorvo-Maki-Nakagawa-Sakata [5]. However, there is available a number of experimental facts which point out necessity of extending this scheme.

The first of them is related to the so-called LSND anomaly [6], which was later investigated in MiniBooNE and MINOS [7,8] experiments. Another experimental evidence for it is Ga anomaly [9,10], that appeared in calibration of Ga neutron detectors. Besides, necessity for introducing an additional neutrino type follows from the analysis of initial nuclear synthesis processes [11] as well as from large scale structures being formed in the Universe [12,13]. Moreover, sterile neutrinos are regarded as being candidates for dark matter in the Universe [13].

Finally, in the early 2011 a reactor antineutrino anomaly was claimed [1,2]. An additional analysis of the data relevant to neutrino production in reactors has shown that the calculated neutrino flux should be increased by 3%. Thus, deficiency of registered events in neutrino experiments has emerged. Some contribution (~0.7%) into this deficiency has been made by neutron lifetime alteration in correspondence with new experimental data [14]. Neutron lifetime has decreased approximately by 1% [15,16], correspondingly efficiency of neutrino detectors has increased by 1%, since efficiency of neutron detectors is formed by cross-section of the reaction of inverse neutron beta-decay. As a result, ratio of the experimentally observed neutrino flux to the predicted one has changed from 0.976±0.024 to 0.943±0.023 [1]. The effect concerned makes up 2.5 of standard deviation. This is not yet sufficient to be sure of existence of the reactor antineutrino anomaly. It should be noted that the effects discussed earlier are also at the validity level 2.5 – 3.0 of standard deviations, thus performance of new and more accurate experiments are considered to be extremely significant.

We have studied opportunity of performing new experiments using research reactors in Russia. The most favorable conditions for carrying out an experiment on search for neutrino oscillations at short distances can be provided by the reactor SM-3. Advantages of the reactor SM-3 are low background conditions, the reactor compact zone of 35x42x42 cm$^3$ at high reactor intensity of 100 MW, as well as fairly small distance of 5 m from the reactor core center to the experimental hall wall. Besides, of great significance is the fact that neutrino flux can be measured within sufficiently wide distance range from 6 to 13 meters. Up to $3 \cdot 10^3$ neutrino events per day can be expected to occur at the reactor intensity of 100 MW at the distance of 6 m from the reactor core in the volume of 1 m$^3$. We have elaborated an experimental project NEUTRINO-4 for 100 MW reactor SM-3 to test the hypothesis of the «reactor antineutrino anomaly» [17].

In order to make preparations for the experiment NEUTRINO-4 at the reactor SM-3 we have carried on experimental investigations at the reactor WWR-M. The task of the experiment is to register neutrino from the reactor in conditions of considerable background from cosmic rays on the Earth surface as well as in conditions of neutron and gamma background in the experimental hall of the research reactor. This experiment was to measure the signal - background ratio at the WWR-M reactor and to investigate basically opportunity of performing such an experiment at the reactor SM-3 by comparing background conditions.

The detector outlay is shown in Fig. 1. The detector with volume of 0.9x0.9x0.5 m$^3$ is divided by a partition into two sections. The partition is applied to prevent light from emitting outside the sections. The detector employs 16 photomultiplier tubes FEU-49B (with each section containing 8) situated on the upper surface. The detector of scintillation type is based on using reaction $\tilde{v}_e + p \rightarrow e^+ + n$. At first the detector registers a positron whose energy is determined by that of antineutrino and 2 annihilation gamma quants, each having energy equal to 511 keV. Neutrons emerging in the reaction are absorbed by Gd to form a cascade of gamma quants, with total energy being about 8 MeV. The detector will register two subsequent signals produced by positron and neutron. The antineutrino spectrum will recover from the positron spectrum, since in the first approximation the correlation between positron energy and that of antineutrino is linear: $E_{\tilde{v}} = E_{e^+} + 1.8$ MeV. Mineral oil (CH$_2$) with addition of Gd 1 g/l is used as material for a scintillator. The BC 525 scintillator light emission is equal to $7 \cdot 10^3$ photons for 1 MeV.

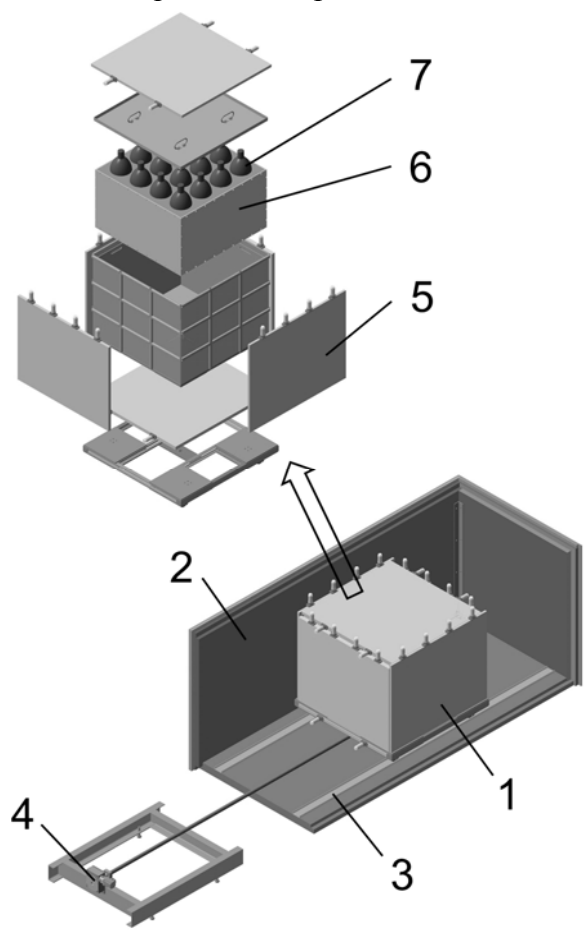

Fig. 1. The scheme of installation model at reactor WWR-M to search for reactor antineutrino oscillations into a sterile state. 1 – reactor antineutrino detector, 2 – passive shielding, 3 – rails, 4 – engine for detector displacement, 5 – scintillator plates of internal anticoincidence shielding with PMT, 6 – volume with scintillator liquid (~400 l), 7 – PMTs of detector.

The detector is surrounded with six scintillation plates 0.9x0.9x0.03 m³ with PMTs which serve as an internal anticoincidence shielding from the cosmic ray muons. The whole construction is surrounded with lead shielding of 7 cm, concrete shielding of 45 cm and another one made of borated polyethylene as thick as 8 cm. The suppression factor of neutron and gamma background at the operating reactor is 4-5 orders of magnitude. On the cover of a passive shielding there is located an external active shielding embracing the area of 8 m². It consists of 32 blocks 0.5x0.5x0.125 m³ of plastic scintillator with PMT. The detector can be displaced inside the shielding for measuring dependence of neutrino flux on the distance, thus embracing the distance range due to its double section structure 5.1 - 6.8 m from the reactor core center to the section center.

Fig. 2a shows the detector energy spectrum. High energy part of the spectrum corresponds to muons having passed through the scintillator in the vertical direction and having the maximum length of path in the scintillator. The edge trajectories form a low energy part of the muon spectrum. The lowest part of the spectrum contains signals from neutron capture by Gd, signals from positrons of neutrino reaction and finally, signals from background gamma quants. The time spectrum or the spectrum of delayed coincidences in Fig. 2b consists of two exponential curves. The first of them with characteristic parameter $\tau_1 \approx 2.2$ μs corresponds to muons which at first produce a starting signal, stop and then decay. Therefore a characteristic parameter of this exponential curve is determined by muon lifetime. A more detailed investigation of this exponential curve shows that it contains 10% contribution of the prompt exponential curve with $\tau_2 \approx 0.31$ μs which is determined by capture of negative muons by nuclei. Finally, the exponential curve with a characteristic parameter $\tau_n \approx 30$ μs is determined by neutron lifetime prior to its capture by Gd at Gd concentration 0.1%.

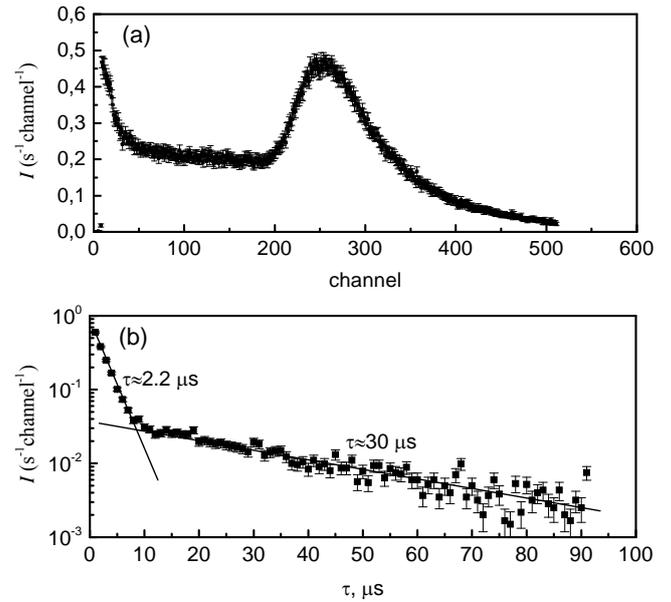

Fig. 2. Detector energy spectrum (a) and time spectrum of delayed coincidences (b). Channel size is 220 keV.

For distinguishing the signal sought for registration of reactor antineutrino one should use both the mode of anticoincidences with muon active shielding and that of delayed coincidences, the scheme of which is given in Fig.3.

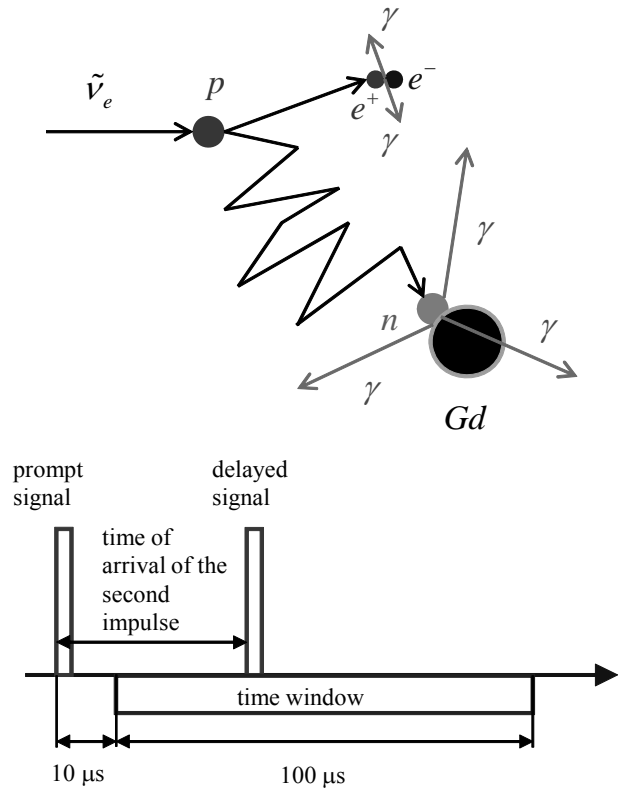

Fig. 3. Process of reactor antineutrino registration and the scheme of delayed coincidences.

Application efficiency of modes shown above is given in Fig. 4. Muon peak is seen to be suppressed practically completely while in the area of searching for a neutrino signal the background is suppressed by three orders of magnitude.

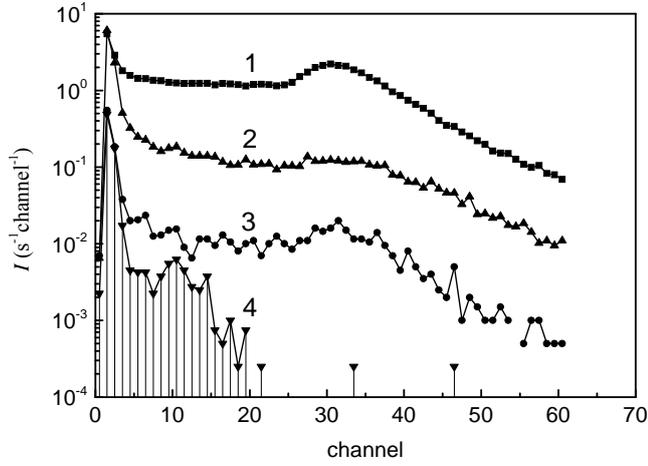

Fig. 4. Curve 1 is energy spectrum of prompt processes in the detector. Curve 2 is an energy spectrum of prompt processes after switching on anticoincidences with internal muon shielding. Curve 3 corresponds to the mode of delayed coincidences. Finally, curve 4 corresponds to the mode of delayed coincidences with switched on internal muon shielding and, in addition, in the delayed coincidences mode the first ten microseconds responsible for muon processes were excluded. Channel size is 1800 keV.

The expected rate of antineutrino events comprises 220 days$^{-1}$ or $2.5 \cdot 10^{-3}$ s$^{-1}$ at distance of 5.3 m from detector center to the reactor core center for 100% detector efficiency. However, the detector count rate within this energy range is almost by the order of magnitude higher and detector efficiency is about 30%. Choosing an optimal energy range (2-8 MeV) for starting and stopping signals, one has succeeded in detecting signal of the reactor ON (the reactor OFF). The time spectrum of delayed coincidences at the reactor ON and the reactor OFF is presented in Fig. 5 for two distances of the detector center from that of the reactor core – 5.3 m and 6.6 m.

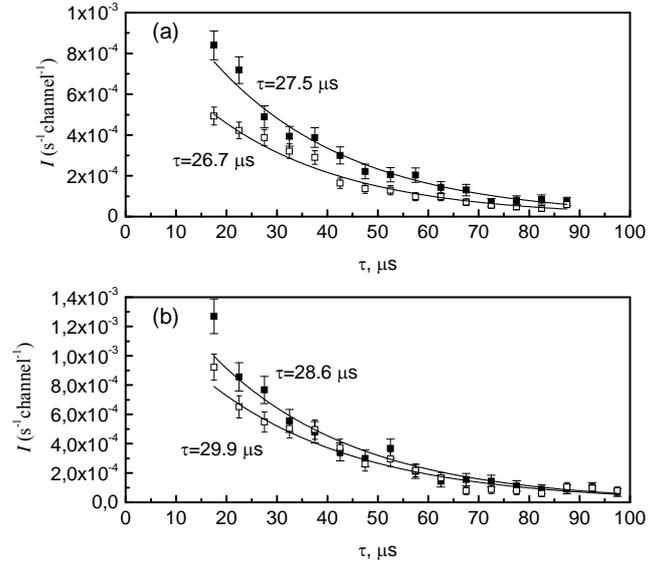

Fig. 5. (a) Time spectrum of delayed coincidences for the position of a detector center 5.3 m from the reactor core center at the reactor ON (■) and at the reactor OFF (□).
(b) Time spectrum of delayed coincidences for the position of a detector center 6.6 m from the reactor core center at the reactor ON (■) and at the reactor OFF (□).

Table 1 shows measuring results at the reactor ON and OFF, their difference and the signal-background ratio. The obtained result for neutrino effect $\Delta_{\text{on-off}} = (1.95 \pm 0.24) \cdot 10^{-3}$ s$^{-1}$ is represented as overestimated one since efficiency of the detector to registration of neutrino events has to be about 30%. It is quite possible that overestimation is connected with existence of fast neutrons from the reactor. Fast neutrons can imitate a neutrino event where a starting signal is the signal from recoil proton of a fast neutron, and stop signal is capture of the slowed-down neutron. Unfortunately, detailed studying of such false effect had not been carried out in connection with a reactor stop. Further studying of false effect from fast neutrons is extremely important. This false effect can be suppressed using the method of discrimination of signals in an impulse form since the signal from recoil proton has to have "long tail" [18].

Table 1. Results of measurements.

| | Reactor ON $I$, $10^{-3}$ s$^{-1}$ | Reactor OFF $I$, $10^{-3}$ s$^{-1}$ | $\Delta$, $10^{-3}$ s$^{-1}$ on-off | $\Delta$/OFF |
|---|---|---|---|---|
| 1+2 section R=5.3m | 6.43 ± 0.20 | 4.48 ± 0.13 | 1.95 ± 0.24 | 0.43 ± 0.055 |
| 1+2 section R=6.6m | 6.09 ± 0.26 | 5.07 ± 0.21 | 1.02 ± 0.34 | 0.21 ± 0.066 |

For drawing a conclusion concerning the possibility of performing an experiment on search for reactor antineutrino oscillations into a sterile state at the reactor SM-3, the comparative measurements of background conditions had been carried out. One of the apparent factors of refining the effect-background ratio is the reactor SM-3 antineutrino intensity being 5.5 higher. The second important factor is the background of fast neutrons of hadron component from cosmic rays. The latter is dependent on the construction of the building, i.e. the amount of conventional and hard concrete located over the neutrino installation. Direct comparative measurements have shown that the flux of fast neutrons in the building designed for a neutrino laboratory at SM-3 (not having any passive shielding for the time being) is $4 \cdot 10^{-4}$ n/cm$^2$s, while at the reactor WWR-M being protected, it is $4 \cdot 10^{-3}$ n/cm$^2$s, i.e. by an order of magnitude higher. In addition, one has made comparative measurements of cosmic ray muon fluxes: 80 m$^{-2}$s$^{-1}$ at the reactor WWR-M and 60 m$^{-2}$s$^{-1}$ at the reactor SM-3. This points out that there is opportunity of obtaining the effect-background ratio being about unit or more at the reactor SM-3. In terms of this effect-background ratio the experiment on search for reactor antineutrino oscillations into a sterile state at the reactor SM-3 seems to be feasible. The detector will contain 5 sections with the total volume of liquid scintillator 1.5 m$^3$. The range for displacements allows making measurements from 6 to 13 m. At present the neutrino laboratory room at the reactor SM-3 has been prepared and the 60-tonn passive shielding made of lead and borated polyethylene is being installed. Detailed description of the project is presented in paper [17].

This work has been carried out at the support of Ministry of Education and Science of the Russian Federation, agreement № 8702. The investigation has been supported by the Russian Foundation for Basic Research, grant № 12-02-12111-ofi_m.